\documentclass[%
reprint,
groupedaddress,
longbibliography,
nofootinbib,
amsmath,amssymb,
floatfix,
aip,
author-year,
]{revtex4-1}

\usepackage[utf8]{inputenc}
\usepackage{amsthm,amsmath,amssymb}

\newcommand{\doOp}{\mathrm{do}}
\newcommand{\E}{\mathrm{E}}
\newcommand{\Var}{\mathrm{Var}}
\newcommand{\Cov}{\mathrm{Cov}}
\newcommand{\citrate}{\lambda}
\newcommand{\latcitrate}{\phi}
\newcommand{\journalcitmult}{\theta}
\newcommand{\ajlatcitrate}{\Phi}
\newcommand{\sjlatcitrate}{\epsilon}

\usepackage{graphicx}
\usepackage[table]{xcolor}

\definecolor{Set1-blue}{RGB}{55,126,184}
\usepackage[colorlinks=true,%
 linkcolor = {Set1-blue},%
 citecolor = {Set1-blue},%
 urlcolor = {Set1-blue}]{hyperref}

\usepackage{orcidlink}

\graphicspath{ {figures/} }

\begin{document}

\title{Inferring the causal effect of journals on citations}

\author{V.A. Traag\,\orcidlink{0000-0003-3170-3879}}
\email{v.a.traag@cwts.leidenuniv.nl}
\affiliation{Centre for Science and Technology Studies (CWTS), Leiden University, the Netherlands}

\date{\today}

\keywords{Science of science $|$ Journal effects $|$ Citations $|$ Causal inference $|$ Bayesian model}

\begin{abstract}
Articles in high-impact journals are, on average, more frequently cited.
But are they cited more often because those articles are somehow more ``citable''?
Or are they cited more often simply because they are published in a high-impact journal?
Although some evidence suggests the latter, the causal relationship is not clear.
We here compare citations of preprints to citations of the published version to uncover the causal mechanism.
We build on an earlier model of citation dynamics to infer the causal effect of journals on citations.
We find that high-impact journals select articles that tend to attract more citations.
At the same time, we find that high-impact journals augment the citation rate of published articles.
Our results yield a deeper understanding of the role of journals in the research system.
The use of journal metrics in research evaluation has been increasingly criticized in recent years and article-level citations are sometimes suggested as an alternative.
Our results show that removing impact factors from evaluation does not negate the influence of journals.
This insight has important implications for changing practices of research evaluation.
\end{abstract}

\maketitle

\section{Introduction}

Journals play a central role in scholarly communication, yet their role is also contested.
The journal impact factor in particular has been criticized on several accounts \citep{Lariviere2018}.
The main critique is its pervasive use in the context of research evaluation, for example in tenure decisions \citep{McKiernan2019}.
Scientists shape their research with impact factors in mind \citep{Rushforth2015,Muller2017}.
In a meeting in San Francisco in 2012, cell biologists called for a ban on the impact factor from research evaluation, and conjoined the ``San Francisco Declaration on Research Assessment''\footnote{https://sfdora.org} (DORA).
A group of researchers and editors called for publishing entire citation distributions instead of impact factors, to counter inappropriate use \citep{Lariviere2016}.
More recently, a group of editors and researchers came together and called for  ``rethinking impact factors''\citep{Wouters2019}.

At the same time, journal impact is one of the most clear predictors of future citations \citep{Callaham2002,Levitt2011,Stegehuis2015,Abramo2019}.
The question is why.
Possibly, high-impact journals select articles that somehow tend to be cited frequently.
Another possibility is that articles are cited more frequently \emph{because} they are published in a high-impact journal, not because they tend to be cited frequently \emph{per se}.
Neither citations of an article nor the journal in which it is published needs to be representative of ``quality''.
Here, we simply study whether citations of an article are influenced by the journal in which it is published, not their relationship to ``quality''.

\begin{figure}[t]
  \centering
  \includegraphics{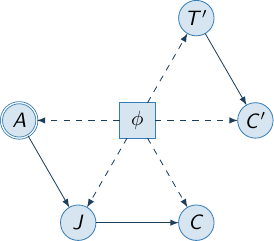}
  \caption{
      Simple causal model of the confounding effect of the latent citation rate $\latcitrate$ of an article being published in a journal $J$ and the citations it accrues $C$.
      In contrast, citations of preprints $C'$ are affected by the latent citation rate $\latcitrate$ only.
      The selection bias on arXiv preprints $A$ does not bias the causal effect of $J$ on $C$ once $\latcitrate$ is controlled for.
      The time before publication $T'$ affects preprint citations $C'$ and complicates the analysis.
      }
  \label{fig:causal_model}
\end{figure}

Answering this question is not straightforward.
In rare cases, publications appear in multiple journals, and researchers found that the version in a higher impact journal was more frequently cited than its twin in a lower impact journal \citep{Lariviere2010,Perneger2010,Cantrill2016}.
However, duplicate publications are quite special, limiting the generalizability of this observation.
Some other earlier work claimed that citations were not affected by the journal \citep{Seglen1994}.

We answer this question by comparing citations of preprints with citations of the published version.
The number of citations $C$ may be influenced by both the latent citation rate $\latcitrate$ and the journal $J$ in which the article is published (Fig.~\ref{fig:causal_model}).
Possibly, high-impact journals perform a stringent peer review of articles, selecting only articles with a high latent citation rate, so that $\latcitrate$ influences the journal $J$.
The latent citation rate itself may be influenced by many factors and characteristics \citep{Onodera2015} and motivations for citing the paper \citep{Bornmann2008}.
These factors are not limited to characteristics of the paper itself, but may also include author reputation \citep{Petersen2014} or institutional reputation \citep{Medoff2006}.
Regardless of which factors influence the latent citation rate, the number of citations of the preprint before it is published in a journal $C'$ is unaffected by where it will be published and is affected only by the latent citation rate $\latcitrate$.
We rely on this insight to estimate the causal effect of the journal on citations $\Pr(C \mid \doOp(J))$.
The identification of the causal effect is possible because of the so-called ``effect restoration'' \citep{Kuroki2014}, provided we can estimate $\Pr(C' \mid \latcitrate)$.
We construct a parametric model that provides exactly such an estimate.

\section{Methodology}

We gathered information about $1\,341\,016$ preprints from arXiv, and identified the published version for $727\,186$ preprints (54\%) (see Appendix~\ref{sec:data} for more details).
We extracted citations of both the preprint version and the published version from references in Scopus.
Preprint dates, publication dates and citation dates are all extracted from Crossref, using a daily granularity.
We used the major subject headings of arXiv as field definitions.
The impact of journals is calculated as the average number of citations received in the first five years after publication for all research articles and reviews in Scopus.
We perform our analysis per year (2000--2016) and field, as the journal effect may vary per year and field.
Moreover, we restrict our analysis to journals that have at least 20 articles that were published at least 30 days after appearing as a preprint on arXiv (Fig.~\ref{ext:fig:preprint_times}).
Clearly, our data has a selection bias \citep{Bareinboim2012} on papers being submitted to arXiv or not ($A$).
However, we can show that this does not affect our estimate of the causal effect $\Pr(C \mid \doOp(J))$ (see Appendix~\ref{app:sec:model}).

Time complicates our analysis.
The time $T'$ before a preprint was published, the \emph{preprint duration}, will clearly affect the number of pre-publication citations $C'$, while the total time since publication $T$ will affect the post-publication citations $C$.
Preprints with a higher latent citation rate may perhaps be more quickly published, thus affecting $T'$.
To tackle this problem, we model the full temporal dynamics of both pre- and post-publication citations.

Citation dynamics are influenced by a wide range of factors, such as a rich-get-richer effect and a clear temporal decay \citep{Fortunato2018}, but was captured reasonably well by a recent model by \citet{Wang2013}.
We build on that model and include a parameter that modulates the citation rate based on where the article is published.
We assume that the number of citations $c_i(t)$ article $i$ receives at time $t$ is distributed as
\begin{equation}
  c_i(t)\sim \mathrm{Poisson}\left[\citrate_i(t) f_i(t)\left(m + C_i(t - 1)\right)\right],
\end{equation}
with effective citation rate $\citrate_i(t)$ and $C_i(t) = \sum_{\tau = 0}^t c_i(\tau)$ the cumulative number of citations, and $m$ a parameter affecting the initial citation accumulation.
The temporal decay of the accumulation of citations is captured by $f_i(t)$, which is modelled by an exponential distribution, with inverse rate $\beta_i$.
We assume that preprint $i$ attracts citations at an effective rate of $\latcitrate_i$, where $\latcitrate_i$ is the latent citation rate of article $i$.
The published version attracts citations at an effective rate of $\latcitrate_i \journalcitmult_{J_i}$, where $\journalcitmult_{J_i}$ is the \emph{journal citation multiplier} for journal $J_i$ in which article $i$ is published.
We equate $\journalcitmult_j$ with the causal effect on citations of publishing in journal $j$, which is  identical for all articles published in journal $j$, regardless of the characteristics of those papers.
We call $C'_i = C_i(T'_i)$ the pre-publication citations and $C_i = C_i(T_i) - C_i(T'_i)$ the post-publication citations.
The expected number of long-term citations is about
\begin{equation}
  m (e^{\latcitrate_i \journalcitmult_{J_i}} - 1),
  \label{equ:approx}
\end{equation}
assuming pre-publication citations are negligible (see Appendix~\ref{sec:model:analysis}).

\begin{figure}[t]
  \centering
  \includegraphics{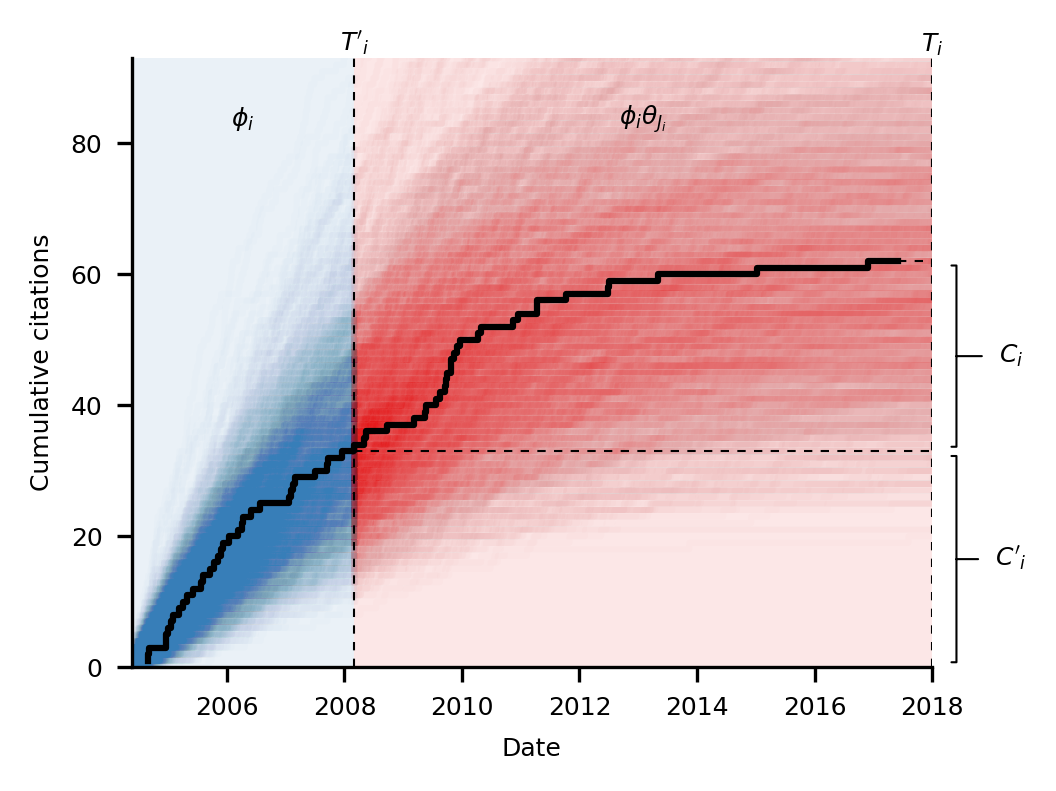}
  \caption{Illustration of citation dynamics.
  This example, \texttt{astro-ph/0405353}, was first submitted to arXiv in 2004 and was published in \emph{Journal of Cosmology and Astroparticle Physics} almost four years later ($T'_i = 1\,385$).
  It was cited 33 times before it was published ($C'_i = 33$), and 29 times after it was published ($C_i = 29$).
  We assume citations are attracted at a rate of $\latcitrate_i$ before it was published and at a rate of $\latcitrate_i \journalcitmult_{J_i}$ after it was published.
  The thick solid line represents the empirically observed number of citations.
  The thin lines in the background represent samples from the posterior predictive distribution of our model.
  }
  \label{fig:example_fit}
\end{figure}

The selection of articles by peer review is assumed to lead to a distribution of latent citation rates for journal $j$,
\begin{equation}
  \latcitrate_i \sim \mathrm{LogNormal}(\ajlatcitrate_j, \sjlatcitrate_j).
\end{equation}
If $\ajlatcitrate_j$ is high, journal $j$ will tend to publish articles of higher latent citation rates $\latcitrate_i$.
The median latent citation rate of journal $j$ is $e^{\ajlatcitrate_j}$.
Effectively, this is a Bayesian hierarchical model, and we specify informed prior distributions based on earlier results \citep{Wang2013} (see Appendix~\ref{sec:model:analysis} for full details and analysis of the model).
We illustrate the model in Fig.~\ref{fig:example_fit}.

\begin{figure*}[t]
  \centering
  \includegraphics[width=\textwidth]{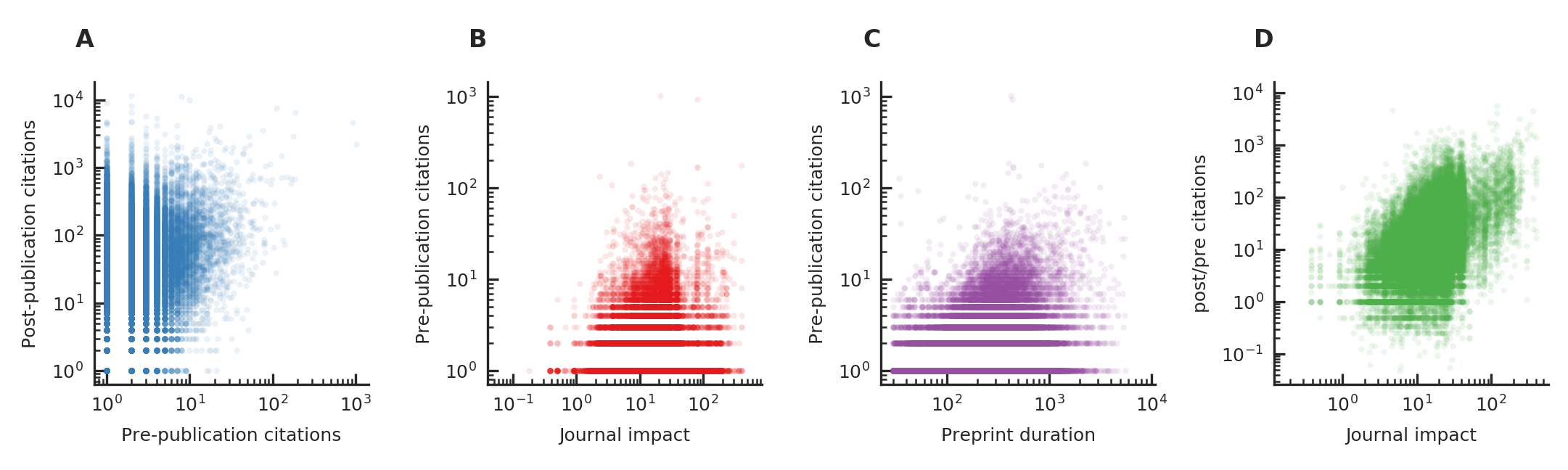}
  \caption{Impact versus pre- and post-publication citations.}
  \label{fig:scatter}
\end{figure*}

\section{Results}

The number of pre- and post-publication citations are not clearly related (Fig.~\ref{fig:scatter}A).
The number of pre-publication citations also do not clearly relate to journal impact (Fig.~\ref{fig:scatter}B).
The relation between preprint duration and the number of pre-publication citations is also not clear (Fig.~\ref{fig:scatter}C).
The ratio of post-publication citations and pre-publication citations is higher for high-impact journals (Fig.~\ref{fig:scatter}D).
Articles in high-impact journals accumulate more post-publication citations relative to pre-publication citations compared to articles that have appeared in lower impact journals.
These results are possibly obfuscated by two counteracting effects: higher latent citation rates lead to higher pre-publication citations, but perhaps also to shorter preprint durations, reducing the time to attract pre-publication citations.
The model that we constructed is intended to address this issue.

We here report results from our model for the five largest fields and the publication year 2016.
Other fields and years show qualitatively similar results (see Figs.~\ref{ext:fig:detailed_results} and~\ref{ext:fig:field_year_overview}).
Our model presents a good fit of both pre- and post-publication citations (Fig.~\ref{ext:fig:citations_fit}).

\begin{figure*}[tb]
  \centering
  \includegraphics[width=\textwidth]{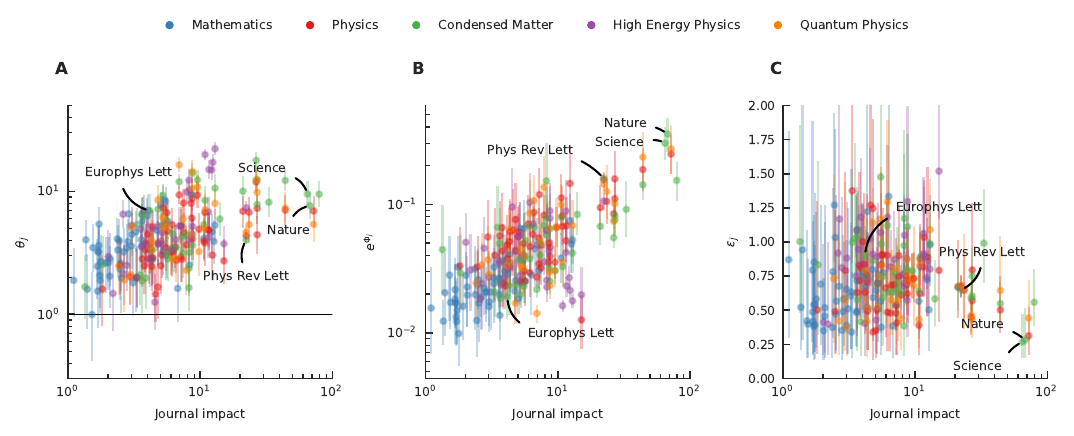}
  \caption{Posterior results for model of citation dynamics for five largest fields and publication year 2016.
    Error bars represent the average $95\%$ credible interval.
    Highlighted journals indicate results in the field of Condensed Matter.}
  \label{fig:results}
\end{figure*}

The journal citation multiplier is consistently higher than $1$ (Fig.~\ref{fig:results}A).
Publishing in journals, compared to being available on arXiv only, multiplies the citation rate substantially, as expected.
For example, \emph{Nature} shows a multiplier of
$6.0$--$9.9$ (95\% CI) for papers published in 2016 in the subject of Condensed Matter and \emph{Science} shows a multiplier of $7.5$--$12.0$ (95\% CI) for such papers.
Using the median estimates and the approximation in Eq.~\ref{equ:approx}, this implies that a Condensed Matter article published in \emph{Nature} in 2016 that obtained about $200$ citations, would have obtained not even $10$ citations had it been available on arXiv only.
Had it been published in \emph{Science} instead, it would have obtained almost $350$ citations.
This is only an illustration: both parameter estimates and the citation dynamics themselves exhibit considerable uncertainty (see Appendix~\ref{sec:model:analysis}).

Most relevant to our question, higher impact journals tend to show higher citation multipliers.
The correlation between the (logarithm of) the journal impact and the (logarithm of) the median journal citation multiplier $\journalcitmult_j$ is on average 0.45 for each combination of field and year.
It ranges from 0.063 for High Energy Physics in 2002 to 0.79 for Astrophysics in 2012.
Interestingly, the correlation grows stronger for High Energy Physics and Astrophysics over time, hovering around 0.6--0.7 for recent years (Fig.~\ref{ext:fig:correlation_dynamics}).

At the same time, the median latent citation rate $e^{\ajlatcitrate_j}$ is also clearly increasing with journal impact (Fig.~\ref{fig:results}B).
For example, the US based \emph{Physical Review Letters} has a relatively high journal impact and shows a latent citation rate of $0.15$--$0.17$ (95\% CI) for Condensed Matter in 2016.
Its lower impact European counterpart \emph{Europhysics Letters} shows a latent citation rate of $0.013$--$0.027$ (95\% CI) in the same field and year.
Overall, the correlation between the (logarithm of) the journal impact and $\ajlatcitrate_j$ is on average 0.54 for each combination of field and year.
For High Energy Physics in 2002 the correlation is 0.72, while for Astrophysics in 2012 the correlation is 0.050.
The highest correlation of 0.85 is observed for Astrophysics in 2006.
This correlation grows weaker for High Energy Physics and Astrophysics over time (Fig.~\ref{ext:fig:correlation_dynamics}).
The median effective citation rate of a journal is $e^{\ajlatcitrate_j} \theta_j$, which aligns closely with the observed journal impact (Fig.~\ref{ext:fig:median_cit_rate}).

The latent citation rates also vary within journals, which is controlled by $\sjlatcitrate_j$.
Journals with a higher $\sjlatcitrate_j$ tend to publish articles with a larger variety of latent citation rates.
For example, \emph{Europhysics Letters} shows a $\sjlatcitrate_j$ of $0.7$--$1.1$ (95\% CI), while \emph{Science} shows a $\sjlatcitrate_j$ of $0.2$--$0.3$ (95\% CI), resulting in a broader distribution of $\latcitrate_i$ for \emph{Europhysics Letters} than \emph{Science}.
In general, high-impact journals show more narrow distributions of latent citation rates than lower impact journals (Fig.~\ref{fig:results}C).

\section{Discussion}
\label{sec:discussion}

Why articles in high-impact journals attract more citations is a fundamental question.
We provided clear evidence that articles in high-impact journals are highly cited because of two effects.
On the one hand, articles that attract more citations are more likely to be published in high-impact journals.
On the other hand, articles in high-impact journals will be cited even more frequently because of the publication venue.
This amplifies the cumulative advantage effect for citations \citep{DeSollaPrice1976}.

A recent publication \citep{Kim2019} took a similar approach and compared citations of preprints with citations of the published version.
Using a more rudimentary model they obtained similar results and also find an influence of the journal on citations, although they do not address the causal mechanism.
They also find that preprints with more citations are more likely to be published, but do not analyse in what journals they are published.

Several mechanisms may play a role in the causal effect of journals on citations.
High-impact journals tend to have a higher circulation \citep{Peritz1995}, and reach a wider audience.
In addition, researchers may prefer to cite an article from a high-impact journal over an article from a low-impact journal, even if both articles would be equally fitting.
Both mechanisms are consistent with our results and earlier results \citep{Lariviere2010,Perneger2010,Cantrill2016,Kim2019}.
Distinguish between these two causal mechanisms is difficult \citep{Davis2010} and should be investigated further.

An alternative explanation may be that published preprints are more highly cited because the preprints were improved by high-quality peer review in high-impact journals.
We deem this an unlikely scenario.
Differences between the preprint and the published version are textually minor \citep{Klein2016}.
Those modifications can of course be substantively important.
Peer review may substantially improve and strengthen a manuscript.
Nonetheless, we think it is unlikely to alter a paper's core contribution so as to affect its citation rate considerably.

Our analysis is limited to mostly physics and mathematics because of our reliance on arXiv.
We expect to see similar effects in the medical sciences and the social sciences, in line with earlier results \citep{Lariviere2010,Perneger2010,Cantrill2016}.
It would be interesting to replicate our analysis on younger preprint repositories, such as bioRxiv or SocArxiv, once they have had more time to accumulate citations.
Another limitation is that we considered references from published articles only.
It would be interesting to include also the references of preprints.
This presumably increases the number of pre-publication citations \citep{Lariviere2014}, which may decrease the overall inferred journal causal effect.

\begin{figure}[b]
  \centering
  \includegraphics{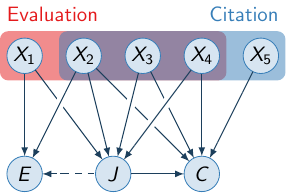}
  \caption{Causal model of factors and characteristics $X_1$, $X_2$, \ldots, journals $J$, citations $C$ and evaluation $E$.}
  \label{fig:complex_causal_model}
\end{figure}

In our model we assumed that the effect of publishing in a journal is identical for all articles published in that journal.
However, the effect of publishing in a journal may possibly vary for different articles.
For example, articles from well-known authors may be cited frequently regardless of the exact journal in which they are published, while articles from more junior authors may benefit more from publishing in high-impact journals.
Teasing out these different effects is not straightforward, but presents an interesting avenue for future research.

The latent citation rate itself may be influenced by many factors and characteristics of the paper \citep{Onodera2015} and motivations for citing the paper \citep{Bornmann2008}.
Overall, our results suggest that characteristics ($X_1$, $X_2$, \ldots) that drive citations ($C$) overlap or correlate with factors that drive journal ($J$) peer review (Fig.~\ref{fig:complex_causal_model}).
For example, novelty, relevance and scientific breadth ($X_2$ to $X_4$) may affect both journal evaluation and citations directly, while methodological aspects affect journal evaluation ($X_1$) and authors' reputation ($X_5$) only affects citations.
Because the journal also affects citations, methodological aspects would have an indirect effect on citations in this example.
What factors drive journal evaluation and what factors drive citations is not clear and should be further investigated.

We hypothesize that a subset of factors that are used in journal evaluation are also used in post-publication research evaluation, such as the UK REF \citep{Traag2019}.
This means that research evaluation ($E$) tends to correlate with journals ($J$) because of underlying common factors (Fig.~\ref{fig:complex_causal_model}).
Even if factors that influence research evaluation do not influence citations directly, they will still correlate because of the mediating effect of the journal.
For example, if methodological aspects ($X_1$) affect research evaluation ($E$), it would correlate with citations ($C$) only because methodological aspects affect the journal ($J$).
If our hypothesis holds true, citations would be indicative of the evaluation of articles only \emph{because} they were published in a particular journal.
In that case, citations should not be normalized based on the journal in which they are published, as was attempted by~\citet{Zitt2005}.
Doing so would effectively control for the journal, thereby blocking these causal pathways.
Indeed~\citet{Adams2008} find that journal-normalized citations do not correlate with evaluation.
Similarly, \citet{Eyre-Walker2013} report an absence of various correlations with evaluations when controlling for the journal.
These results provide some evidence for our hypothesis.
Journal metrics might even be a more appropriate indicator than citations to individual articles, as was suggested by \citet{Waltman2017}, although our results neither affirm nor refute this possibility.

Possibly, evaluation itself is also affected directly by the journal in which an article is published, and depending on the context, perhaps also by its citations.
Indeed, the proposed causal diagram only captures part of a larger web of entanglement.

The use of citations and journals in research evaluation is often debated.
Removing the use of journal metrics from research evaluation, as for example advocated by DORA, may decrease the pressure on authors to publish in high-impact journals.
The use of article-level citations for evaluation could be condoned by DORA, but the use of journal metrics could not.
Even if journal metrics were to be removed from research evaluation, journals would continue to play a role in research evaluation, albeit indirectly.
Evaluating researchers based on citations then may still reward authors who publish in high-impact journals.
This may effectively exert selective pressures that drive the evolution of the research system \citep{Smaldino2016}.
Simply removing impact factors from research evaluation therefore does not negate the influence of journals.

\begin{acknowledgments}
  \noindent I thank Rodrigo Costas, Ludo Waltman, Jesper Schneider and other colleagues from CWTS.
  I gratefully acknowledge use of the Shark cluster of the LUMC for computation time.
\end{acknowledgments}

\section*{Conflict of interest}

\noindent The author declares no conflict of interest.

\section*{Data availability}

All data necessary to reproduce the results in this analysis is available from \citet{data} and all source code is available from \citet{code}.

\bibliography{bibliography}

\onecolumngrid
\clearpage

\appendix
\renewcommand{\thefigure}{S\arabic{figure}}
\setcounter{figure}{0}

\section{Data}
\label{sec:data}

We combined data from arXiv, Crossref and Scopus to establish our dataset.
All data necessary to reproduce the results in this analysis is available from \citet{data} and all source code is available from \citet{code}.

\subsection{arXiv}
\label{sec:data:arxiv}

We downloaded data from a bulk export from arXiv from \\
https://archive.org/download/arxiv-bulk-metadata and used the file arxiv\_biblio\_oai\_dc.2018-01-19.xml.

For all arXiv XML elements in the data we extracted the arXiv identifier, and if present the DOI.
We also extracted the date the preprint was first posted on arXiv.
In total, this dataset covered $1\,341\,016$ preprints, and a DOI is provided for $727\,186$ preprints (54\%).

We extracted the subject for each arXiv preprint.
The subjects were quite noisy, and did not contain only the subject division of arXiv, but also other subject classifications, most notably, the Mathematical Subject Classification (MSC).
The arXiv subject classifications were provided as ``Major - Minor'' subjects, although sometimes only a major subject was provided.
We extracted the major part and assigned an arXiv preprint to a major subject if that subject is at least used by $1\,000$ preprints (and is not an MSC).
We thus retain 18 major subjects.

Preprints can be assigned to multiple major subjects.
The large majority of arXiv preprints is assigned to a single major subject (80\%).
A single preprint has been assigned to as many as $8$ different major subjects (\texttt{1108.2700}).
There are only $261$ preprints that have not been assigned to any of the major subjects.
These are papers that are published in economics (33) and electrical engineering (228), subjects which were introduced in September 2017, and in which arXiv did not yet have many preprints at the time of data collection.

\subsection{Crossref}
\label{sec:data:crossref}

We established the publication date using Crossref, which is available in-house at CWTS.
We used the Crossref database that was imported on August 2018.
We determined the publication date as the first date of the following dates from Crossref: ``published online'', ``published print'', ``created'' and ``issued.
We established the publication date for all arXiv preprints.
Out of the $727\,186$ provided DOIs in arXiv, we find a match in Crossref for $722\,003$ articles (99\%).

We established the publication date for all citing publications using Crossref in the same way.
See the next paragraph for more details concerning the citing publications.

\subsection{Scopus}
\label{sec:data:scopus}

The Scopus database is available in-house at CWTS, which we used for our analysis.
We relied on the Scopus database that was imported on May 2018.

We used Scopus to find the published version of the preprint.
This was done by matching the DOI from arXiv with the DOI as recorded in Scopus.
Out of the $722\,003$ DOIs from arXiv that were matched to Crossref, we found $664\,741$ DOIs from Scopus with a unique match (92\%).
We used the matched publication in Scopus to identify the journal in which the preprint was published.

We calculated the impact of journals using Scopus.
We defined the impact as the average number of citations received in the first five years after publication for all articles (document type \texttt{ar}) and reviews (document type \texttt{re}).
For articles that were published within five years of the end of the database (2018), we counted
citations until the end of the database.

Finally, we used Scopus to identify citations of both the preprint version and the published version.
We parsed all raw cited reference strings provided in Scopus to extract an arXiv identifier or a DOI.
We identified arXiv identifiers in the reference string using the regular expression
  \begin{verbatim}
      [a-zA-Z\-\.]+ ?/ ?[0-9]{7,}|[aA][rR][xX][iI][vV]:
      [0-1][0-9]([0][0-9]|[1][0-2])\.[0-9]{4,5}
  \end{verbatim}
If the reference was matched by Scopus, and a cited publication was identified, we used the DOI from the cited publication as recorded in Scopus.
If that was not available, we used the DOI in the reference string extracted using the regular expression
  \begin{verbatim}
    \b10\.[0-9]{4,}(\.[0-9]+)*/\S*\b
  \end{verbatim}
We identified $4\,679\,896$ references with arXiv identifiers in more than half a billion references in total.

For all citing documents, we extracted the publication date through Crossref, as described earlier.
We used this date as the cited date of the cited document.
The cited date is used at the resolution of a day.
Note that some reference may still cite the preprint, even if the preprint is published, although most citations after the preprint is published refer to the published version, as already observed earlier by \citet{Lariviere2014}.
For clarity, we define citations that were made on or before the publication date of the preprint as pre-publication citations, and we define citations that were made after the publication date as post-publication citations.
In total we identified $156\,528$ pre-publication citations and $15\,939\,887$ post-publication citations from references in Scopus.

\section{Model and Bayesian inference}
\label{app:sec:model}

There is a clear selection bias \citep{Bareinboim2012} on papers being submitted to the arXiv or not ($A$).
We assume that the latent citation rate $\latcitrate$ may affect whether a paper will be submitted to the arXiv $A$, which in turn may affect the journal $J$.
Previous research showed that publications that are available as preprints are more highly cited \citep{Lariviere2014,Fu2019}, but this ``citation advantage'' seemed unlikely to be causal \citep{Davis2008,Gaule2011}.
We therefore assume the arXiv does not directly influence the citations $C$.
If we control for $\latcitrate$ (which is effectively done by controlling for $C'$), we obtain that $\Pr(C \mid \doOp(J), A = 1, \latcitrate) = \Pr(C \mid \doOp(J), \latcitrate)$ by the rules of do-calculus \citep{Pearl2009}.
We thus obtain an unbiased estimate of the causal effect $\Pr(C \mid \doOp(J))$, even if our observations are biased towards arXiv papers, as stated in the main text.

\begin{figure}
  \centering
  \includegraphics{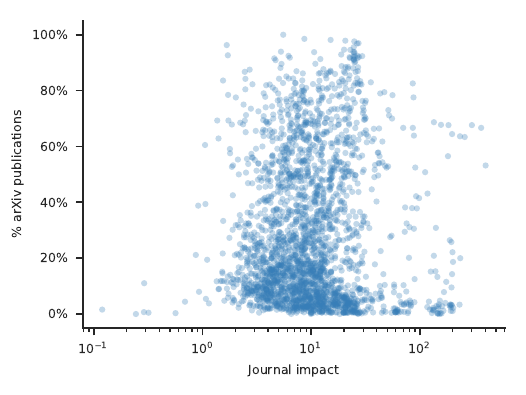}
  \caption{Percentage of publications that are available as preprints on arXiv. This is limited to only preprints that have been posted on arXiv at least 30 days before publication.}
  \label{ext:fig:arxiv_source_published}
\end{figure}

As explained previously, the journal causal effect $\Pr(C \mid \doOp(J))$ is not affected by the selection on arXiv papers $A$.
The same does not hold for the estimate $\Pr(J \mid \doOp(\latcitrate))$, as $A$ could possibly act as a mediator.
Possibly, authors decide to to only post preprints they deem sufficiently good.
Being posted on the arXiv may possibly affect where it is could be published, for example, because some journals may have policies against publishing preprints.
In our causal diagram, $\latcitrate$ may then affect $A$ which in turn may affect $J$.
Because of the selection effect on $A$, the effect of $\latcitrate$ on $J$ then perhaps only holds for arXiv preprints.
To better understand this possible mediating effect, we computed for each journal the proportion of arXiv papers it published.
We only included arXiv papers that had at least a preprint duration of at least 30 days.
We find there is no discernible relationship between journal impact and the proportion of arXiv papers~(Fig.~\ref{ext:fig:arxiv_source_published}).
In other words, $A$ is unlikely to act as a mediator, suggesting that high-impact journals indeed select articles with higher latent citation rates.
Although this observation is again confounded by the latent citation rate $\phi$, it would be rather surprising to have a confounding effect that exactly cancels out the actual causal effect of $A$ on $J$, so that we observe no correlation between $A$ and $J$.

The full specification of the hierarchical Bayesian model introduced in the main text is as follows.
As already introduced in the main text, we model the probability of attracting $c_i(t)$ citations at time $t$ as
\begin{equation}
  c_i(t) \sim \mathrm{Poisson}\left(\citrate_i(t) f_i(t)\left(m + C_i(t - 1)\right)\right)\\
  \label{equ:cit_distribution}
\end{equation}
with $m$ some parameter affecting the initial rate of attracting citations and
\begin{equation}
  \citrate_i(t) = \left\{
    \begin{array}{rl}%
      \latcitrate_i & t \leq T'_i \\
      \latcitrate_i \journalcitmult_{J_i} & t > T'_i
    \end{array}
  \right..
\end{equation}
where $T'_i$ is the date at which publication $i$ is published in a journal with $t = 0$ the time at which the preprint was posted on arXiv.
We are modelling citations at a daily rate, and it is reasonable to assume that citations on the same day have not influenced each other.
Citations on the same day can be regarded as independent events.
The Poisson distribution models exactly a random variable that counts the number of evens that happen at a given rate within a given interval, making it a suitable distribution for $c_i(t)$.
This is a slight generalization from the earlier model by \citet{Wang2013} who only consider the probability of being cited at a certain time $t$.
In practice, publications may attract multiple citations at a single day, and we therefore consider the number of citations explicitly.
This happens only infrequently, as only about $6\%$ of the days at which a publication is cited is it cited more than once in our dataset.

The temporal decay is represented by $f_i(t)$, which follows the density of an exponential distribution
\begin{align}
  f_i(t) &= \int_{t}^{t + 1} \frac{1}{\beta} \exp \left[ -\frac{\tau}{\beta} \right] d\tau \\
         &= \exp \left[ -\frac{t}{\beta} \right] - \exp \left[ -\frac{t+1}{\beta} \right]
\end{align}
We define $F_i(t) = \sum_{\tau = 0}^t f_i(t)$, so that
\begin{equation}
  F_i(t) = 1 - \exp \left[ -\frac{t+1}{\beta} \right].
\end{equation}
For the temporal decay we assume a prior of
\begin{equation}
  \beta_i \sim \mathrm{InvGamma}(2, 3 \times 365).
  \label{equ:prior:decay}
\end{equation}
Our prior expectation is that the decay takes about $3$ years, which corresponds roughly to earlier results \citep{Wang2013}.
This agrees also with other literature on the decay of citations \citep{Egghe1992,Avramescu1979,Parolo2015}.
Note that we do not use the log-normal distribution for the decay, as used in the model on which we build \citep{Wang2013}.
Modelling the decay using the log-normal distribution resulted in problem of convergence, which seemed to be due to multimodality of the logarithmic decay, problematizing model identifiability.
Using a maximum likelihood approach as used in the earlier work \citep{Wang2013} may miss this multimodality.
Using an exponential decay improved the convergence of the Bayesian sampling.
Note that even an exponential decay can lead to an initial increase of the number of citations and later decrease, as is typical of citations.
We show this in when analyzing the model in more detail in the next section.

There is a certain degeneracy in the model for pre-publication citations that depends on our assumptions of the prior for the decay.
If we observe few pre-publication citations, this can be due to two factors: a low decay $f_i(t)$ at that point $t$, or a low $\latcitrate_i$.
It is therefore important to assume reasonable priors for the temporal decay.
If we assumed that $f_i(t)$ would be mostly concentrated in the first few days, we would erroneously infer a too low $\latcitrate_i$ and a too high $\journalcitmult_{J_i}$.
Although an exponential decay by definition only decreases, our prior expectation is that the decay is quite gradual.
The prior on $\beta_i$ is also quite broad, allowing for substantially different decay.

We assume that the latent citation rate of articles published in a certain journal $j$ is distributed as
\begin{equation}
  \latcitrate_i \sim \mathrm{LogNormal}(\ajlatcitrate_j, \sjlatcitrate_j).
\end{equation}
We assume priors of
\begin{align}
  \ajlatcitrate_j &\sim \mathrm{Normal}(0, 1), \\
  \sjlatcitrate_j &\sim \mathrm{InvGamma}(2, 1).
\end{align}
which roughly corresponds to distributions of $\citrate_i$ as found in \citep{Wang2013} for various journals, assuming the journal citation multiplier is about $1$.
Although $\latcitrate_i$ is modelled hierarchically as an element of a journal, causally speaking, $\latcitrate$ determines $J$, not the other way around.
That is, there is certain causal effect $\Pr(J \mid \doOp(\latcitrate))$, which we assume to give rise to the probability $\Pr(\latcitrate | J)$ we model here.
The use of priors in fitting this type of models is also employed in a response \citep{Wang2014a} to some critique of the model \citep{Wang2014}.
In line with \citep{Wang2014a} we simply set $m=30$ and do not infer this parameter from the data.

Finally, we assume the following prior on the journal citation multiplier $\journalcitmult_j$
\begin{equation}
  \journalcitmult_j \sim \mathrm{Gamma}(2, 2),
\end{equation}
which is centered around $1$.

The larger citation rates observed for high-impact journals may correspond to either a higher $\ajlatcitrate_j$ or a higher $\journalcitmult_j$.
Our priors are relatively conservative with respect to a journal causal effect.
We have assumed a prior on $\ajlatcitrate_j$ that corresponds to overall distribution citation rates as found in earlier work \citep{Wang2013}.
The prior on $\journalcitmult_j$ is centered around $1$, corresponding to no journal causal effect, but still allows for larger $\journalcitmult_j$.

We use \texttt{pystan} \texttt{2.19.0} to perform Bayesian inference of the posterior distributions using the no-U-turn sampler \citep{StanDevelopmentTeam2017}.
In practice, citations are relatively sparsely distributed throughout time and  $c_i(t) = 0$ for most $t$.
Instead of specifying the probability for each $t$ separately, we can more efficiently specify the probability for only those $t$ for which $c_i(t) > 0$.
The probability of observing $0$ citations for a duration of $\tau$ is identical to an exponential distribution with the same rate as the Poisson distribution in Eq.~\ref{equ:cit_distribution}.
More specifically, for a $t_1$ and $t_2$ such that $c_i(t_1) > 0$ and $c_i(t_2) > 0$, the probability of observing $0$ citations for all $t$ between $t_1$ and $t_2$ then equals
\begin{equation}
  \Pr(C_i(t_2 - 1) - C_i(t_1) = 0) = \exp \left[ - \citrate_i(t_1) (m + C_i(t_1)) (F_i(t_2 - 1) - F_i(t_1)) \right]
\end{equation}
assuming times $t_1$ and $t_2$ do not cross the publication date $T'_i$.
In they do cross $T'_i$, the time windows $(t_1, T'_i]$ and $(T'_i, t_2)$ should be considered separately.
To improve the numerical stability of \texttt{pystan}, we use a logarithmic specification of the rate for the Poisson distribution.
This also necessitates to work with the logarithm of the temporal decay, which has a simple form.
Finally, we use four chains of $1\,000$ iterations each, using half of the iterations for warmup with a target acceptance rate of $0.98$ (\texttt{adapt\_delta}) and a maximum tree depth of $20$.

\begin{figure*}[bh]
  \centering
  \includegraphics[width=\textwidth]{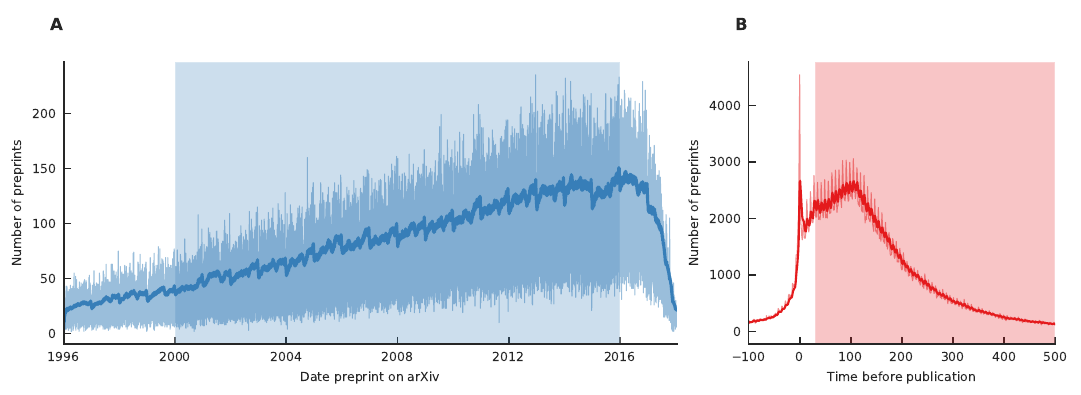}
  \caption{Preprints on arXiv. (a) the number of preprints submitted to arXiv per day; (b) the time before a preprint is published ($T'_i$). The shaded areas indicate what part of the data is used for estimating the journal causal effect.}
  \label{ext:fig:preprint_times}
\end{figure*}

We perform our analysis per year (2000--2016) and field, and restrict to journals that have at least $20$ articles that were published at least 30 days after being posted as a preprint on arXiv (Fig.~\ref{ext:fig:preprint_times}).
This results in $3\,892$ different subsets that are separately fitted.
The different subsets cover $258$ different journals.
There were seven subsets which yielded diverging transitions.
Only one subset showed large problems, and almost $25\%$ of the transitions diverged.
The remaining six subsets only showed three diverging transitions at most.
Nonetheless, we excluded all subsets that showed diverging transitions, but results are unaffected by the exclusion or inclusion of these seven problematic subsets.
Using log-normal temporal decay resulted in diverging transitions for about two-third of the subsets.

Source code for fitting our model is available in the Zenodo repository \\
https://doi.org/10.5281/zenodo.3583012.

\subsection{Analysis}
\label{sec:model:analysis}

We first analyse the mean number of citations attracted by article $i$.
We can write the total number of citations $C_i$ as $C_i(t) = C_i(t-1) + c_i(t)$ for $t > 0$ with $C_i(0) = c_i(0)$.
Taking the expected value then yields
\begin{equation}
  \E(C_i(t)) = \E(C_i(t-1)) + \E(c_i(t)).
\end{equation}
Writing out the expected number of citations received at time $t$ yields
\begin{align*}
  \E(c_i(t)) &= \sum_{C=0}^\infty \E(c_i(t) \mid C_i(t - 1) = C) \Pr(C_i(t - 1) = C) \\
             &= \sum_{C=0}^\infty \citrate_i(t) f_i(t) (m + C) \Pr(C_i(t - 1) = C) \\
             &= \citrate_i(t) f_i(t) (m + \E(C_i(t - 1))),
\end{align*}
so that we end up with the recursion
\begin{equation}
  \E(C_i(t)) = \E(C_i(t - 1)) + \citrate_i(t) f_i(t)\bigl(m + \E(C_i(t - 1))\bigr).
  \label{eq:recursion_mean}
\end{equation}
This recursion has as a solution
\begin{equation}
  \E(C_i(t)) = m \left(\prod_{\tau = 0}^{t} (1 + \citrate_i(\tau) f_i(\tau)) - 1 \right),
\end{equation}
which can be easily checked by substituting in Eq.~(\ref{eq:recursion_mean}):
\begin{align*}
  \E(C_i(t)) &= \E(C_i(t-1)) + \citrate_i(t) f_i(t) (m + \E(C_i(t-1))) \\
   &= m \left(\prod_{\tau = 0}^{t-1} (1 + \citrate_i(\tau)f_i(\tau)) - 1 \right)  + \citrate_i(t) f_i(t) \left(m + m \left(\prod_{\tau = 0}^{t-1} (1 + \citrate_i(\tau)f_i(\tau)) - 1 \right) \right) \\
   &= m \left(\prod_{\tau = 0}^{t-1} (1 + \citrate_i(\tau)f_i(\tau)) - 1 + \citrate_i(t)f_i(t) \prod_{\tau = 0}^{t-1} (1 + \citrate_i(\tau)f_i(\tau)) \right) \\
   &= m \left(\prod_{\tau = 0}^{t} (1 + \citrate_i(\tau)f_i(\tau)) - 1 \right).
\end{align*}
Writing the product as an exponential sum of logarithms we obtain
\begin{equation}
  \E(C_i(t)) = m \left(\exp \left[ \sum_{\tau = 0}^t \log(1 + \citrate_i(\tau)f_i(\tau))\right] - 1 \right).
\end{equation}
A simple Taylor expansion shows that $\log(1 + x) \approx x$ for small $x$, so that we obtain the approximation
\begin{equation}
  \E(C_i(t)) \approx m \left(\exp \left[\sum_{\tau = 0}^t \citrate_i(\tau)f_i(\tau)\right] - 1 \right).
  \label{eq:mean_approximation}
\end{equation}
Expanding $\citrate_i(\tau)$ we obtain
\begin{equation}
  \E(C_i(t)) \approx
  \left\{
    \begin{array}{rl}
      m \left(\exp \left[\latcitrate_i F_i(t)\right] - 1 \right)
        & \mathrm{for~} t \leq T'_i \\
      m \left(\exp \left[\latcitrate_i F_i(T'_i) + \latcitrate_i \journalcitmult_{J_i} (F_i(t) - F_i(T'_i))\right] - 1 \right)
        & \mathrm{for~} t > T'_i
    \end{array}
  \right..
\end{equation}
The expected number of pre-publication citations is given by $\E(C_i) = \E(C_i(T'_i))$ while the expected number of post-publication citations is given by $\E(C_i) = \E(C_i(T_i)) - \E(C_i(T'_i))$ so that we obtain respectively
\begin{equation}
  \E(C'_i) \approx m \left(\exp \left[\latcitrate_i F_i(T'_i)\right] - 1 \right)
\end{equation}
and,
\begin{align}
  \E(C_i) &\approx m \left(\exp \left[\latcitrate_i F_i(T'_i) + \latcitrate_i \journalcitmult_{J_i} (F_i(T_i) - F_i(T'_i)) \right] - 1 \right) - m \left(\exp \left[\latcitrate_i F_i(T'_i)\right] - 1 \right) \\
  &= m \exp \left[ \latcitrate_i F_i(T'_i) \right] \left( \exp \left[ \latcitrate_i \journalcitmult_{J_i} (F_i(T_i) - F_i(T'_i))\right] - 1 \right).
\end{align}
Taking the limit of $t \to \infty$ and assuming pre-publication duration is negligible, we obtain the approximation of the expected number of long-term citations of $m (e^{\latcitrate_i \journalcitmult_{J_i}} - 1)$.

Using the approximation for the total number of citations $E(C_i(t))$ we can also obtain an approximation for the expected instantaneous number of citations.
This approximation shows that the number of citations can initially increase, even if the temporal decay is exponential.
We use a continuous time approximation, and take the derivative of Eq.~\ref{eq:mean_approximation} with respect to $t$ and assume $\theta = 1$ for simplicity.
We then obtain the approximation that
\begin{equation}
  \E(c_i(t)) \approx \frac{m\latcitrate_i}{\beta} \exp \left[ \latcitrate_i \left(1 - e^{-\frac{t}{\beta_i}} \right)-\frac{t}{\beta_i} \right],
\end{equation}
which attains its maximum at $t = \beta_i \log \latcitrate_i$ for $\latcitrate_i > 1$.
This shows that citations first increase and then decrease, similar to what is observed empirically.
Publications with a slower decay attain this peak later.
Similarly, publications that have a higher latent citation rate also attain the maximum at a later time.
Interestingly, this is formally equivalent to an older result \citep{Avramescu1979}.

We can also analyse the variance of $C_i(t)$ and obtain the recursion
\begin{equation}
  \Var(C_i(t)) = \Var(C_i(t - 1)) + \Var(c_i(t)) + 2 \Cov(C_i(t - 1), c_i(t)).
\end{equation}
Since $\Cov(C_i(t - 1), c_i(t)) > 0$ this recursion yields a variance $\Var(C_i(t))$ that is larger than the expected value.
Hence, there is considerable uncertainty in citations in this model, even for an exact $\latcitrate_i$ and $\journalcitmult_j$.
This means that even for specific $\latcitrate_i$ and $\journalcitmult_j$, the distribution of citations would be quite skewed.
It is therefore possible that skewed citation distributions within a journal emerge, even if latent citation rates $\latcitrate_i$ are homogeneously distributed \citep{Waltman2017}.

This result is mostly due to the rich-get-richer effect, also known as Matthew effect or cumulative advantage, which is frequently argued to explain the high variance and skewness observed in most citation distributions, dating back to early literature in scientometrics \citep{DeSollaPrice1976}.
Without the rich-get-richer effect, citations $C_i(t)$ would simply be Poisson distributed around $\sum_{\tau=0}^t \citrate_i(t) f_i(t) m $ according to this model.
In that case, citation distributions tend to be less skewed for specific $\latcitrate_i$, so that the skewness in citation distributions may require a more heterogeneous distribution of $\latcitrate_i$.
We cannot distinguish between these two alternative possibilities based on our empirical observations.
It would be interesting to empirically substantiate the cumulative advantage effect for citations, but this goes beyond the scope of this paper.
In line with previous literature, we assume the presence of a cumulative advantage effect in our model.

\begin{figure}
  \centering
  \includegraphics{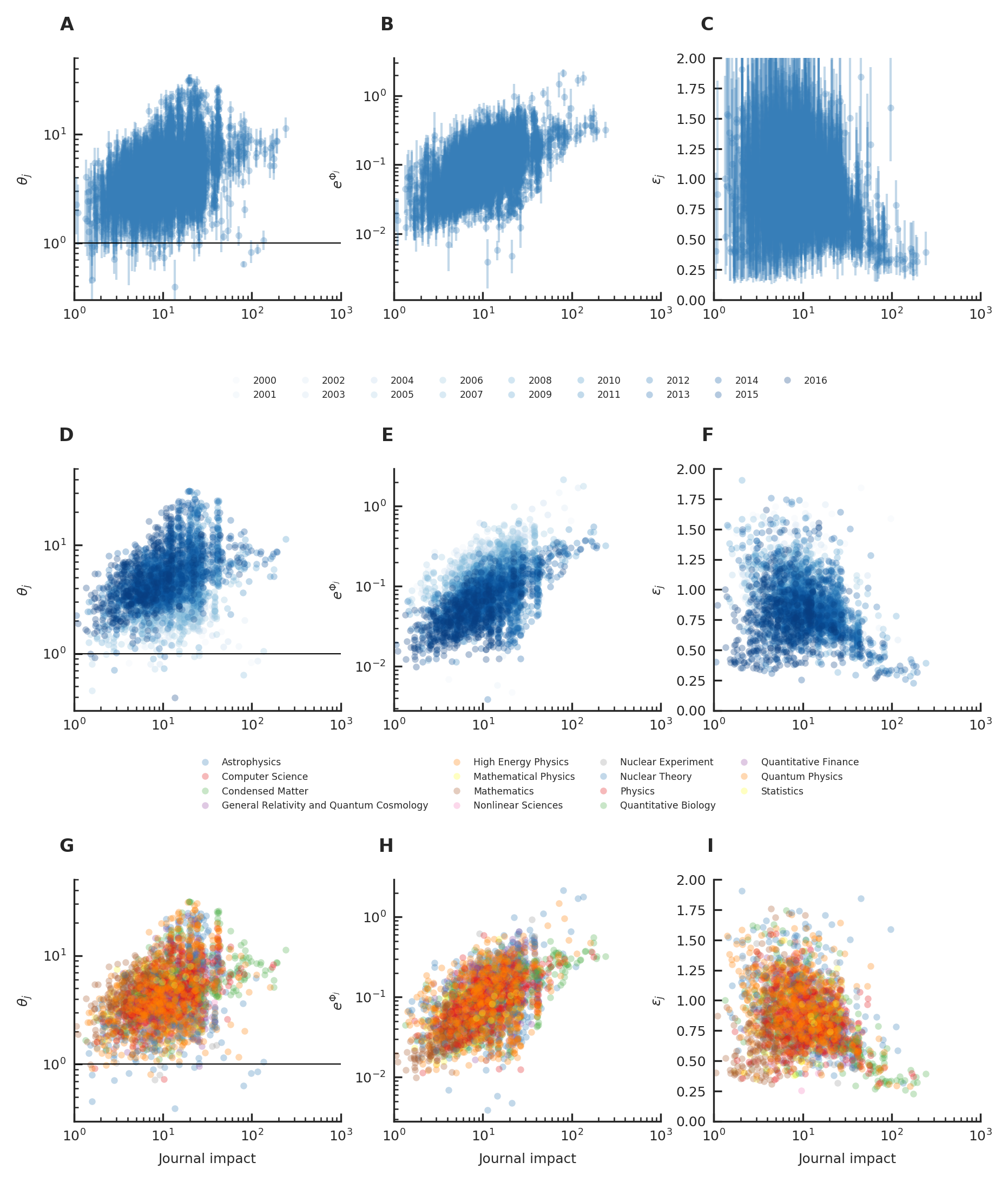}
  \caption{Detailed results.
      This shows the dependency of the citation multiplier $\journalcitmult_j$, the median latent citation rate $e^{\ajlatcitrate_j}$ and the $\sjlatcitrate_j$ on journal impact (a-c).
      The visualization shows the median and the error bars represent the 95\% credible interval.
      This also shows the same results but separated per year (d-f) and field (g-i).
    }
  \label{ext:fig:detailed_results}
\end{figure}

\begin{figure}
  \centering
  \includegraphics{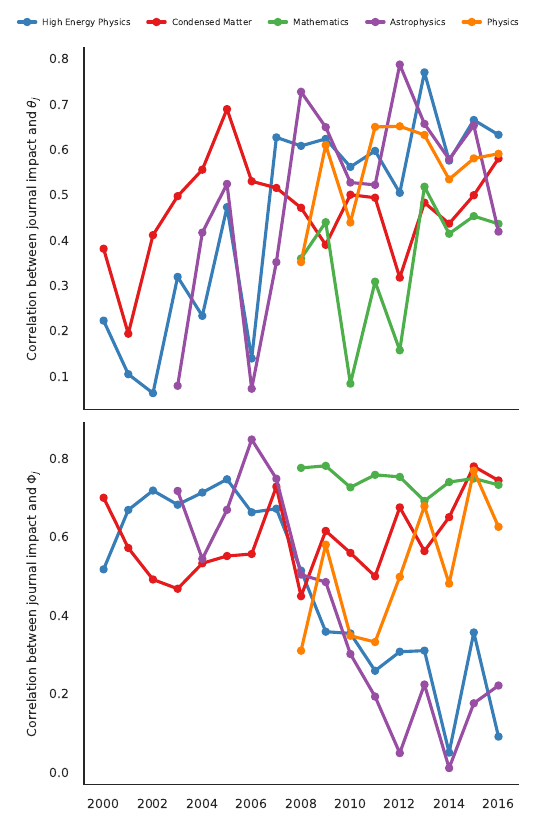}
  \caption{Correlation dynamics.
      The correlation is provided for the five largest fields over time, for years that have at least 20 journals present.
      We use the median estimate for the journal citation multiplier $\journalcitmult_j$ and the latent citation rate $\ajlatcitrate_j$ to calculate correlations.
      We take the logarithm of the journal impact, and the logarithm of the median journal citation multiplier $\journalcitmult_j$ before calculating the correlations.
      The correlation between the journal impact and the journal citation multiplier $\journalcitmult_j$ seems increasing for High Enery Physics and Astrophysics over time (a), while the correlation between journal impact and latent citation rate $\ajlatcitrate_j$ is decreasing over time (b).
    }
  \label{ext:fig:correlation_dynamics}
\end{figure}

\begin{figure}
  \centering
  \includegraphics{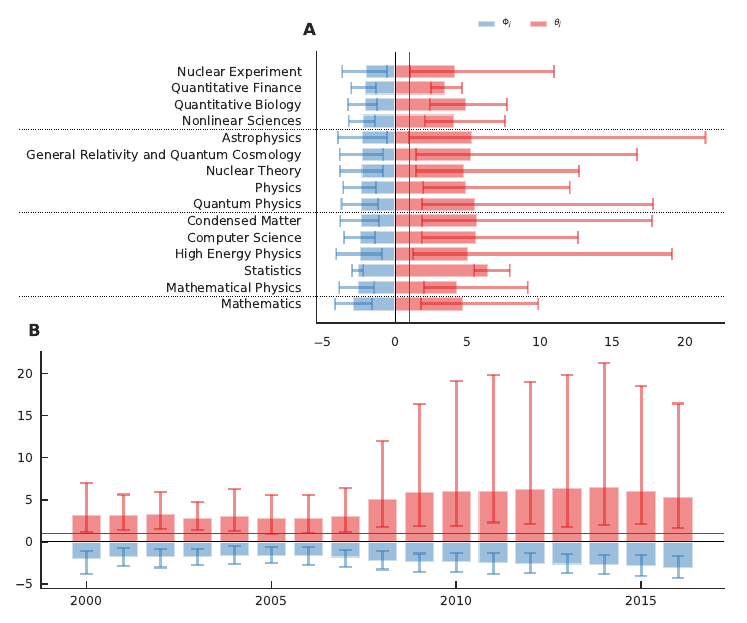}
  \caption{Overview per field and year.
    Distribution of median estimates of $\ajlatcitrate_j$ and $\journalcitmult_j$ for (a) field and (b) year.
    Error bars indicate 95\% percentile intervals of median estimates for journals in specified field or year.
    There is some variation over fields.
    The multiplier $\journalcitmult_j$ seems to be relatively high for Statistics, whereas Quantitative Finance shows a relatively low multiplier.
    Possibly, statisticians do not regularly follow new preprints on arXiv.
    There seems to be some trend over the years of increasing journal citation multipliers but the trend is not very clear.
    }
  \label{ext:fig:field_year_overview}
\end{figure}

\begin{figure}[t]
  \centering
  \includegraphics[width=\columnwidth]{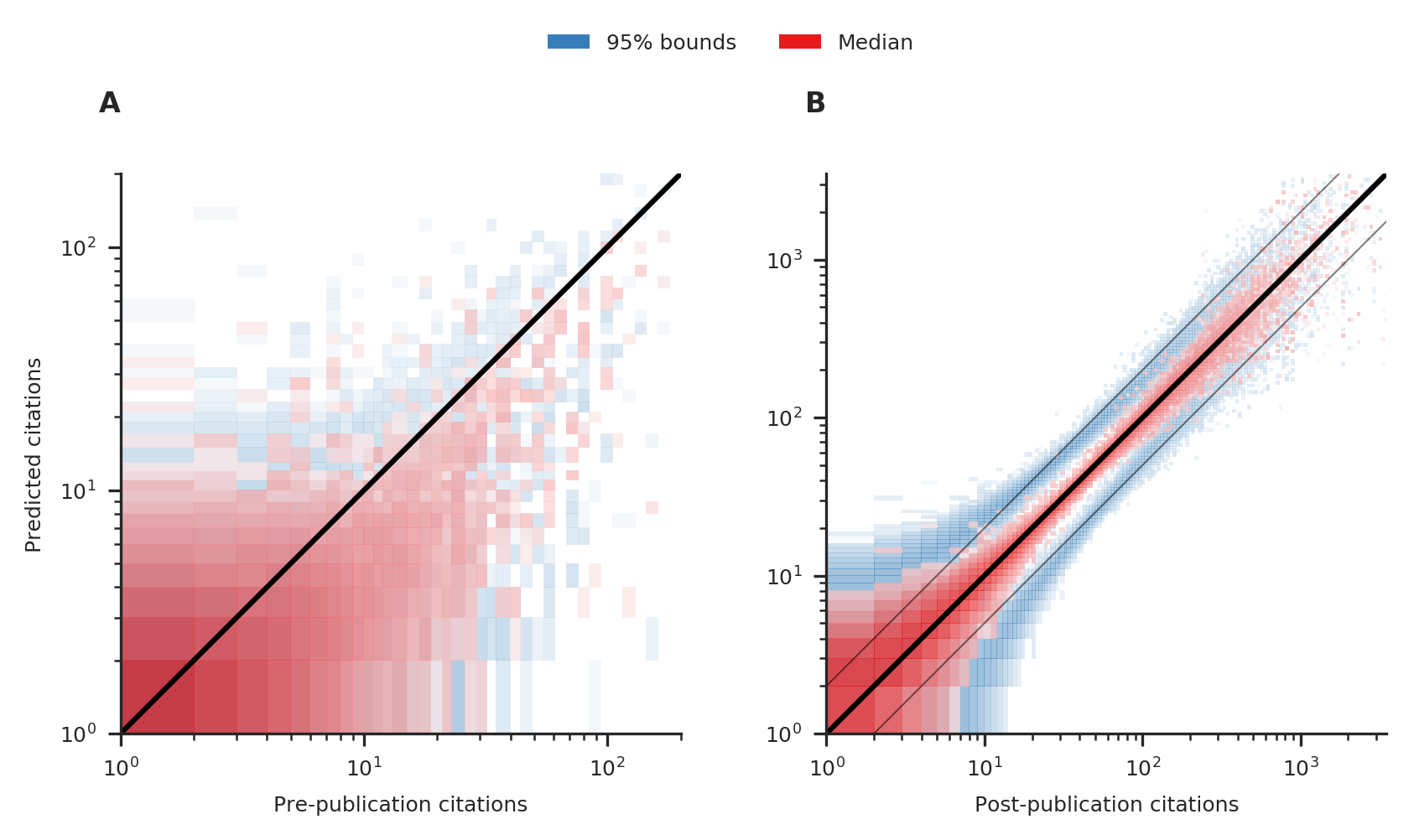}
  \caption{Predicted citations versus observed citations.
  The 95\% credible interval of the predicted number of citations is roughly between half and twice the median predicted number of citations.
  This quantifies both the uncertainty of the inferred parameters as well as the uncertainty arising from the citation dynamics themselves.
  For lower number of citations the credible interval is a bit broader.}
  \label{ext:fig:citations_fit}
\end{figure}

\begin{figure}
  \centering
  \includegraphics{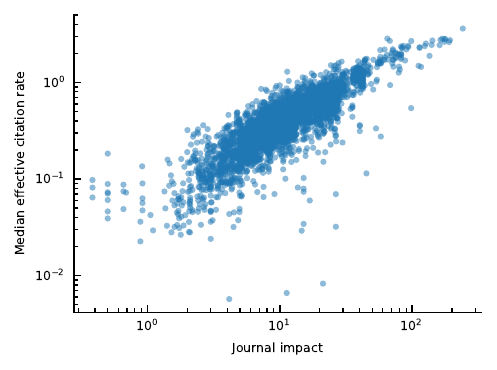}
  \caption{
    Median effective citation rates and journal impact.
    }
  \label{ext:fig:median_cit_rate}
\end{figure}

\end{document}